\documentclass[a4paper,showpacs,prb,twocolumn]{revtex4}
\usepackage{bbm}
\usepackage{amsfonts}
\usepackage{graphicx}
\usepackage{color}
\usepackage{amsmath}
\usepackage{amssymb}
\usepackage{latexsym}
\usepackage{psfrag}

\begin{document}

\title{Edge disorder and localization regimes in bilayer
graphene nanoribbons}
\author{Hengyi Xu}
\email{Hengyi.Xu@uni-duesseldorf.de}
\author{T. Heinzel}
\affiliation{Condensed Matter Physics Laboratory, Heinrich-Heine-Universit\"at,
Universit\"atsstr.1, 40225 D\"usseldorf, Germany}
\author{I. V. Zozoulenko}
\email{Igor.Zozoulenko@itn.liu.se}
\affiliation{Solid State Electronics, Department of Science and Technology, Link\"{o}ping
University, 60174 Norrk\"{o}ping, Sweden}
\date{\today }

\begin{abstract}
A theoretical study of the magnetoelectronic properties of zigzag and
armchair bilayer graphene nanoribbons (BGNs) is presented. Using the recursive
Green's function method, we study the band structure of BGNs in uniform
perpendicular magnetic fields and discuss the zero-temperature conductance for
the corresponding clean systems. The conductance quantized as
$2(n+1)G_{0}$ for the zigzag edges and $nG_{0}$ for the armchair edges with
$G_{0}=2e^{2}/h$ being the conductance unit and $n$ an integer. Special attention is paid to
 the effects of edge disorder. As in the case of monolayer graphene nanoribbons
(GNR), a small degree of edge disorder is already sufficient to induce a
transport gap around the neutrality point. We further perform comparative
studies of the transport gap $E_{g}$ and the localization length $\xi $ in
bilayer and monolayer nanoribbons. While for the GNRs
$E_{g}^{GNR}\sim 1/W$, the corresponding transport gap $E_{g}^{BGN}$ for the
bilayer ribbons shows a more rapid decrease as the ribbon width
$W$ is increased. We also demonstrate that the evolution of localization lengths with the Fermi energy shows two
distinct regimes. Inside the transport gap, $\xi $ is essentially
independent on energy and the states in the BGNs are
significantly less localized than those in the corresponding GNRs. Outside the transport gap $\xi $ grows rapidly as the Fermi energy increases and becomes very similar  for BGNs and GNRs.
\end{abstract}

 \maketitle

\section{Introduction}

The recent successful fabrication of monolayer graphene \cite
{Novoselov} has ignited tremendous interest because it not only represents a
platform to model relativistic particles in a condensed matter material, but
also provides a potential building block for future nanoelectronics. The
honeycomb lattice of the graphene sheet imposes a linear
low-energy electronic spectrum and the corresponding extraordinary electronic
behavior of the excitations near the Dirac point, namely massless Dirac
fermions \cite{Katsnelson,Neto2009}. Various aspects of graphene like band
structures \cite{Wakabayashi1999,Brey2006} and resulting transport properties
\cite{Rojas2006,Lewenkopf2008,Chen2007,Han2007,Wang2008} have been studied.
Electrostatic potentials \cite{Katsnelson} and magnetic barriers
\cite{Martino2007,Xu2008,Masir2008} have been suggested as ways to achieve
tunable confinement. Moreover, there have been
suggestions to induce a bandgap and manipulate efficiently the resistance by,
for example, chemical doping \cite{Filho2007}, edge disorder
\cite{Evaldsson2008}, while studies of the magnetoelectronic properties also
revealed the anomalous integer and fractional quantum Hall effects in graphene
\cite{Zhang2005,Novoselov2005,Gusynin2005,Peres2006}.

\begin{figure}[tbp]
\includegraphics[width=8cm]{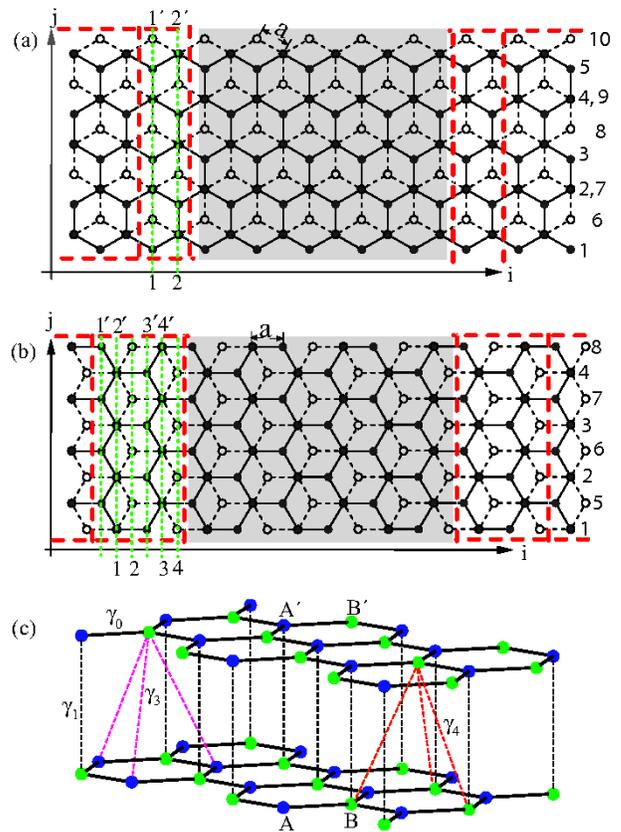} 
\caption{(Color online) Schematic geometry of the two-terminal structure for
the case of (a) zigzag and (b) armchair BGNs. The devices under
consideration are indicated by the shaded regions which connect to the two
semi-infinite leads opening to the left and right. Bonds in the bottom layer
are indicated by dashed lines and in the top layer by solid lines. Empty and
filled circles denote the atoms in the lower and upper layers, respectively.
Unit cells of the BGNs are marked by bold dashed rectangles. Vertical dashed
lines with numbers label the slices of a unit cell; the numbers with and
without prime labels refer to respective upper and lower layers, for
example, $1$ and $1^{\prime }$ correspond to the first slice. $a $ is the
carbon-carbon bond length. Numbers in the vertical direction label the sites
in a slice. (c) 3D view of the graphene bilayer with all coupling energies.
Two interpenetrating triangular sublattices sublattices in the lower and
upper layers are denoted $A$, $B$, and $A^{\prime }$, $B^{\prime }$,
respectively. $\protect\gamma_0$ is the coupling between nearest-neighboring
sites in the same layer. Interlayer coupling energies include $\protect\gamma%
_1$ between $A^{\prime }$ and $B$, $\protect\gamma_3$ between $B^{\prime }$
and $A$, and $\protect\gamma_4$ between $A^{\prime }(B^{\prime })$ and $A(B)$
($\protect\gamma_4$ is neglected in our calculations).}
\label{fig1}
\end{figure}

More recently, increasing attention has been paid to graphene multilayers, in
particular to bilayer systems \cite
{Mccann2006,Kechedzhi2007,Snyman2007,Abergel2007,Lai2008} which consist of two
coupled graphene sheets with two sublattices, $A$ and $B$, in each layer. They
are typically stacked in the Bernal form, where $A^{\prime }$ sites belonging
to the upper layer are located exactly on the top of $B$ sites to the lower
layer, and $B^{\prime }$ or $A$ sites are above or below the center of hexagons
in the other layer, as shown in Fig. \ref{fig1}. Bilayers show quite different
properties than monolayers in many respects, such as the quantum Hall effect
\cite{Mccann2006,Novoselov2005}, edge states \cite{Castro2008}, and weak
localization \cite{Gorbachev}. Graphene bilayers are also inevitably influenced
by the disorder present in the environment. The role of disorder in graphene
bilayers has been studied theoretically \cite{Nilsson2008} and the minimal
conductivity has been addressed by different authors
\cite{Katsnelson2006,Cserti}. An important property of graphene bilayers is
that an application of an electric field between the layers allows to open up
and tune a band gap, which is highly relevant for the possible applications in
nanoelectronics. Experimentally, a double-gate configuration can impose a
perpendicular electric field onto a bilayer, and a controlled transition from a
zero-gap semiconductor to an insulator has been observed, which provides a
direct evidence of opening a bias-induced band gap in a bilayer
\cite{Oostinga}. It has been shown that a bias voltage can continuously tune a
bandgap from zero to mid-infrared frequencies around the zero-energy point
\cite{Castro2007}, modify the charge density distribution in the graphene
planes \cite{Mccannprb2006} and even induce the low-density ferromagnetism
\cite{Castroprl2008}.

To study the properties of bulk bilayer graphene theoretically, the effective
Dirac-like Hamiltonian as the continuum limit of the tight-binding model close
to the Dirac points was mostly used. It has been applied to the study of the
energy spectrum of bilayers \cite{Falko2009}, to the transport phenomenology
through biased \cite{Nilsson2007} and unbiased \cite{Nilsson2006} graphene
bilayers with bulk disorder as well as to the bilayer/monolayer interface
\cite{Nilsson_prb_2007}. However, the energy spectra and the transport
properties of bilayer graphene nanoribbons (BGNs), which form the focus of this paper, have not yet been addressed in the literature to the best of our knowledge.

The purpose of the present work is twofold. First, we develop a computational
method and present a theoretical study of the magneto-band structure, Bloch
states and conductance quantization in zigzag and armchair BGNs. Our method is
based on the recursive Green's function technique that we recently developed to
calculate the electronic structure and conductance of graphene nanoribbons
\cite{Xu2008}. The special feature of this technique is that, in contrast to
earlier Green's function methods, it does not require self-consistent
calculations for the surface Green's function, which is
instead computed as a solution of an eigenequation. This greatly reduces
the computation time. Making use of this technique, one can study the band
structure and conductance of BGNs of various geometries and edge terminations
in various transport scenarios including a perpendicular magnetic field, a bias
voltage, etc. It should also be noted that the electron-electron interaction
and screening can significantly affect and modify electronic properties of
single-layer graphene nanoribbons \cite{Fernandez-Rossier,Efetov}. It is
expected that screening in the BGNs differs from that one in GNRs, because the electrons in each layer are affected not only by
the electrons belonging to the same layer, but also by the electrons in the
second layer. A detailed knowledge of the electronic structure of an ideal BGNs
(without interaction) is a prerequisite for any studies of interaction effects
in this system. While accounting for electron interactions will be the subject of  future studies, knowledge of electronic structure of the ideal BGN
represents a necessary first step in this direction.

Second, the effects of edge
disorder on the conductance of BGNs are investigated.  It is well established that the edge
disorder strongly affects the transport properties of monolayer graphene
nanoribbons leading to the suppression of the conductance quantization and
opening of the transport gap \cite
{Chen2007,Han2007,Areshkin,Louis,Gunlycke,Li,Querlioz,Avouris,Evaldsson2008,Mucciolo}.
We are not aware of any work addressing the effect of edge disorder on the
transport properties of BGNs and our study represents a first step in this
direction. We demonstrate that conductance of realistic BGNs with edge disorder
share many similar features with that one of the monolayer GNRs. There are
however important and interesting differences that we discuss in detail.

The paper is organized as follows. In Section II, we sketch the geometry of
the devices and the Green's functions technique that we use. In Section III,
we present the band structures and the conductance of ideal BGNs. Section IV
focuses on the effects of edge disorder. A summary and conclusions
constitute Section V.

\section{Model descriptions}

We shall consider the two-terminal BGN structure sketched in Fig. \ref{fig1}%
(a) and (b). The semi-infinite left and right leads are made of zigzag or
armchair BGNs. In the shaded central region, one can define a structure of
interest, such as an electrostatic barrier, point defects, or edge roughness.
In our approach we describe the bilayer by the tight-binding Hamiltonian
\cite{Neto2009}
\begin{eqnarray}
H &=&\sum_{\ell ,\langle i,j\rangle }(V_{\ell ,i}a_{\ell ,i}^{+}a_{\ell
,i}+V_{\ell ,j}b_{\ell ,j}^{+}b_{\ell ,j})-\gamma _{0}\sum_{\ell ,\langle
i,j\rangle }(a_{\ell ,i}^{+}b_{\ell ,j}+h.c.)  \notag \\
&-&\gamma _{1}\sum_{i}(a_{1,i}^{+}b_{2,i}+h.c.)-\gamma _{3}\sum_{\langle
i,j\rangle }(b_{1,i}^{+}a_{2,j}+h.c.)
\end{eqnarray}%
where $a_{\ell ,i}^{+}$ ($b_{\ell ,i}^{+}$) is the creation operator at
sublattice $A$ ($B$), in the layer $\ell =1,2$, at site $\mathbf{R}_{i}$. $%
V_{\ell ,i}$ is the on-site electrostatic potential. We use the common graphite
nomenclature for the coupling parameters \cite{Neto2009} (see illustration in
Fig. \ref{fig1} (c)): $\gamma _{0}=3.16$ $\mathrm{eV}$ is
the intralayer nearest-neighbor coupling energy, $\gamma _{1}=0.39$ $\mathrm{%
eV}$ is the coupling energy between sublattice $B$ and $A^{\prime }$ in
different graphene layer, and $\gamma _{3}=0.315$ $\mathrm{eV}$ the hopping
energy between sublattice $A$ and $B^{\prime }$ in the lower and upper
layers, respectively. The other coupling energy between the
nearest-neighboring layers, $\gamma _{4}\approx 0.04$ $\mathrm{eV}$, is very
small compared with $\gamma _{0}$ and ignored below. In the presence of an
external perpendicular magnetic field $B$, the hopping integral acquires the
Peierls phase factor such that the coupling $\gamma _{0(3)}$ is modified to $%
\gamma _{0(3)}\exp {(ie\theta _{r,r^{\prime }}/\hbar )}$ with $\theta
_{r,r^{\prime }}=\int_{r}^{r^{\prime }}\mathbf{A}\cdot d\mathbf{l}$ and $%
\gamma _{1}$ leaves unchanged, where the Landau gauge, $\mathbf{A}=(-By,0,0)$
was used.

In the Green's function method, it is convenient to write the Hamiltonian $H$
in the form
\begin{equation}
H=\sum_{i}[h_{i}]+U,  \label{H_U}
\end{equation}%
where $h_{i}$ describes the Hamiltonian of the $i$-th slice, and $U$
describes hopping between all neighboring slices. Each $i$-th slice of the
BGN consists of two slices of the length $N$ belonging respectively to the
upper and the lower layers, such that the dimension of the matrixes $h_{i}$
and $U$ is $2N\times 2N.$ The choices of slices are indicated in the Figs. %
\ref{fig1}(a) and \ref{fig1}(b). The numbers label the indices of slices in
the lower layer, while the numbers with primes label the indices of slices
in the upper layer. In the same slice of zigzag BGNs, the transverse sites
of the upper layer are sitting above the sites of the lower layer. For
armchair BGNs, we choose the slices in which the transverse sites of the
lower layer are shifted by a distance of $a/2$ with respect to the sites
belonging to the upper layer in order to include all the coupling between
the nearest-neighboring slices. The Hamiltonian matrix of the $i$-th slice
of the BGN consists of two $N\times N$ sub-matrices describing the slices of
the upper and lower layers, which constitute the diagonal blocks. The
off-diagonal blocks of the Hamiltonian matrix account for the interactions
between the slices of the upper and lower layers. The diagonal blocks of the
hopping matrixes $U$ describe the intralayer hopping between the
nearest-neighboring slices, while the off-diagonal blocks describe the
hopping between the nearest-neighboring slices in the different layers.
Explicit expressions for the matrices $[h_{i}]$ and $U$ are given in the
Appendix \ref{appdx1}.

To calculate the band structures, we first study an infinite ideal BGN by
using the Green's function formalism. A unit cell of the BGN structure
consists of $M$ slices, with $M=2$ for the zigzag and $M=4$ for the armchair
as shown in Figs. \ref{fig1} (a),(b). The Bloch states and the wave
functions of BGNs are obtained by solving an eigenvalue problem which is
formulated in the Appendix \ref{appdx2}. The wave functions of the slices
are further used to calculate the surface Green's functions accounting for
the effects of semi-infinite leads.

In order to calculate the transmission coefficient we divide the structure
into three regions. The left and right leads are modeled by ideal
semi-infinite BGNs, which differ only in that one extends to the left and
the other to the right. These two leads are then connected to one other via
the central device which is indicated by the shaded region in Figs. \ref%
{fig1} (a),(b) (where scattering occurs). The transmission and reflection
amplitudes are related to the total Green's functions which are computed using
the standard recursive technique based on the Dyson equation
\cite{Ferry,Rahachou}. The relevant equations are formulated in the Appendix
\ref{appdx2} and their detailed derivations (for the case of a monolayer
graphene nanoribbons) can be found in Ref. [\onlinecite{Xu2008}]. The
zero-temperature conductance is then calculated using the Landauer-B\"{u}%
ttiker formalism,

\begin{equation}
G = \frac{2e^2}{\hbar}\sum_{\alpha,\beta}(T)_{\beta\alpha},
\end{equation}
where $(T)_{\beta\alpha}$ is the transmission from incoming state $\alpha$
in the left lead to outgoing state $\beta$ in the right lead.

\section{Magneto-band structure and conductance quantization of ideal
bilayer nanoribbons}

\subsection{Magneto-band structure}

In this section we calculate the band structures of BGNs for different edges
as well as in perpendicular magnetic fields, characterized by the magnetic
flux $\phi $ through a unit cell of the graphene lattice in units of the
flux quantum $\phi _{0}=h/e$.

\begin{figure}[tbp]
\includegraphics[width=8cm]{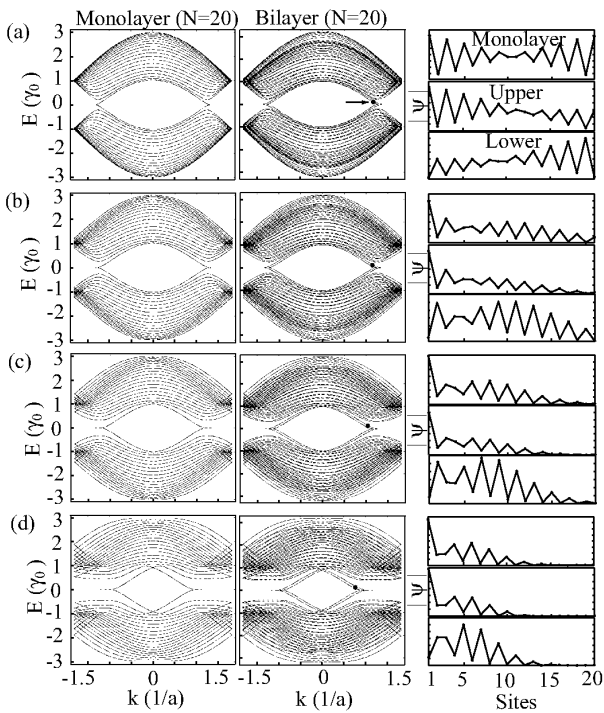} 
\caption{The energy band structures of zigzag monolayer (left panel) and
bilayer (right panel) ribbons with $N=20$ for different magnetic flux
through a hexagon (a) $\protect\phi/\protect\phi_0=0$, (b) $\protect\phi/%
\protect\phi_0=1/200$, (c) $\protect\phi/\protect\phi_0=1/100$ and (d) $%
\protect\phi/\protect\phi_0=1/50$. The corresponding wave functions at the
energy represented by full circles are shown to the right (the top figure
for each $\protect\phi/\protect\phi_0$ shows the wave function of the
monolayer, while below, the wave functions of the upper and the lower layer
are shown).}
\label{fig2}
\end{figure}

\begin{figure}[tbp]
\includegraphics[width=8cm]{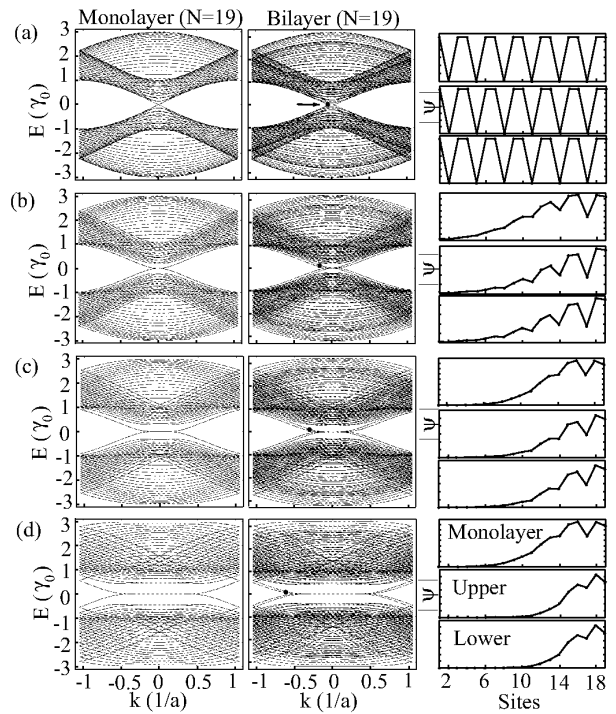} 
\caption{The energy band structures of metallic armchair monolayer (left
panel) and bilayer (right panel) ribbons with $N=19$ for (a) $\protect\phi/%
\protect\phi_0=0$, (b) $\protect\phi/\protect\phi_0=1/200$, (c) $\protect\phi%
/\protect\phi_0=1/100$ and (d) $\protect\phi/\protect\phi_0=1/50$. The
corresponding wave functions are shown to the right. }
\label{fig3}
\end{figure}

In Figs. \ref{fig2} and \ref{fig3} we present the energy band structures of
respectively zigzag and armchair BGNs with $N=20,19$\ and the corresponding
wave functions for the values of magnetic flux $\phi /\phi
_{0}=0,1/200,1/100,1/50$. Dispersion relations of the corresponding monolayer
GNRs are plotted for comparison. As expected, the band structure of BGNs
corresponds to two superimposed and somewhat deformed band structures of
individual monolayer GNRs. To understand this character,
we first assume that the two graphene layers of a bilayer system are decoupled. The system is therefore expected to behave as two isolated
monolayer GNRs with all bands in the dispersion relation being doubly
degenerate. By turning on the coupling between the two layers the degeneracy of
the states is then lifted by the presence of the coupling, with the degree of
splitting depending on the coupling strength. This is in a close analogy to a
textbook example of a coupled quantum wells, where the splitting of the quantum
well states depends on the strength of the barriers separating the wells. For
the case of BGNs the coupling between layers is given by the hopping integrals
$\gamma _{1}$\ and $\gamma _{3}$\ whose magnitudes ($\sim 0.3-0.4$\ $eV$)
determine the value of splitting of the corresponding monolayer bands.

The features of the band structure of BGNs at $B=0$\ outlined above persist
in the presence of a magnetic field. As $B$ increases to $\sim \phi /\phi
_{0}=1/200$\ the bands close to the Dirac points begin to flatten. This
represents the onset of Landau level formation when the classical
cyclotron radius $r_{c}$\ for the low-velocity states (i.e. for those close
to the Dirac points) becomes smaller than the ribbon width ( $r_{c}=\hbar
k/eB=kl_{B}^{2},$\ $l_{B}=\sqrt{\hbar /eB}$\ (=1.3 nm at $\phi /\phi
_{0}=1/200$) being the magnetic length). As the field is increased to $\phi
/\phi _{0}=1/50$, the cyclotron motion dominates and the Landau levels
become well developed.

It is interesting to note that similar to the case of monolayer GNRs, partly
flat bands at $E=0$ exist in the band structures of the zigzag BNGs
corresponding to the localized states on each edge, see Fig. \ref{fig2}.
Compared to the monolayer cases, the number of edge states is doubled for each
edge in bilayer ribbons. A detailed theoretical treatment of edge states in
bilayer systems has been made recently \cite{Castro2008}.

Figures \ref{fig2},\ref{fig3} also show some representative wave functions for
different magnetic fields. As expected, the wave functions in the BGNs closely
follow the corresponding wave functions of the monolayer GNRs. As the magnetic
field increases, the wave functions are pushed towards the ribbon edges,
signaling the formation of the familiar magnetic edge states which correspond
to classical skipping orbits \cite{Ferry}.

It should be noted that our nanoribbons are rather narrow and therefore the
onset of the formation of the Landau levels (when the magnetic length $l_{D}$%
\ becomes comparable to the ribbon's width) can be reached only for an
unrealistically high magnetic fields (for example, $\phi /\phi _{0}=1/50$\
corresponds to $B=400$\ T). In realistic experimental samples of submicron
width, the formation of Landau levels scales down to much smaller magnetic
fields. Due to computational limitations, however, it is rather impractical
to consider wider ribbons and in the present study we therefore consider
the same physics at higher fields.

\begin{figure}[tbp]
\includegraphics[width=8cm]{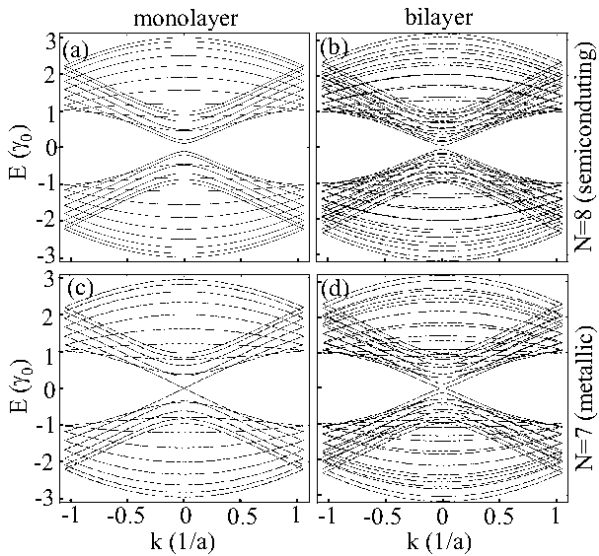} 
\caption{The energy bands of semiconducting and metallic armchair
nanoribbons in $B=0$: (a) monolayer with $N=8$; (b) bilayer with $N=8$. In
(c) and (d), the band structures of the monolayer and bilayer systems with $%
N=7$ are shown.}
\label{fig4}
\end{figure}

The electronic properties of armchair graphene nanoribbons depend
sensitively on the ribbon width. The GNRs are metallic when $N-1$ is a
multiple of $3$; otherwise, there is an energy gap between the conduction
and valence band. Fig. \ref{fig4}(a-b) displays the band structures of the
armchair BGN and GNR at $B=0$ with $N=8,7$. It is evident that both the
monolayer and bilayer graphene ribbons with $N=8$ possess a bandgap around
the zero-energy point. However, the energy gap of the BGN is narrower than
that of the monolayer GNR even though they have the identical width. For the
armchair BGN and GNR with $N=7$, the band structures show metallic
characteristics as shown in Fig. \ref{fig4}(c) and (d). In contrast to the
linear dispersion relation near the Dirac point in the monolayer GNR (Fig. %
\ref{fig4}(c)), the armchair BGN (Fig. \ref{fig4}(d)) shows a parabolic
dispersion relation near the zero-energy point, which leads to new elementary
excitations called massive Dirac fermions. These massive non-relativistic
particles are chiral and can be described by an off-diagonal, Dirac-like
Hamiltonian as the continuum limit of the tight-binding model
\cite{Mccann2006}. This Hamiltonian of the bulk bilayer system gives rise to
two parallel parabolic bands separated by an amount of $\sim \gamma _{1}$ close
to the zero-energy point, which is consistent with our observations for
nanoribbon structures.

\subsection{Conductance quantization}

\begin{figure}[tbp]
\includegraphics[width=8cm]{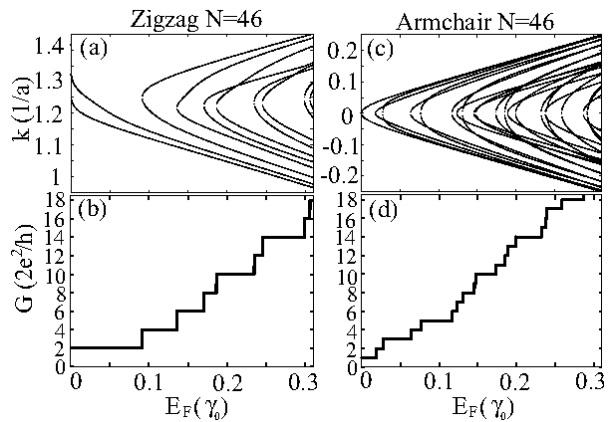} 
\caption{Upper panels: the energy spectra for the zigzag (a) and armchair
(c) bilayer ribbons with $N=46$. (For the zigzag ribbon only the branches in
the vicinity of one Dirac point are shown). Lower panels: The energy
dependence of the zero-temperature conductance of the ideal zigzag (b) and
armchair (d) ribbons associated with the energy spectra.}
\label{fig5}
\end{figure}

Now we are in the position to study the transport properties of BGNs. In clean
ideal nanoribbons the conductance is determined by the number of propagating
modes at the Fermi energy. Each spin-degenerate propagating mode contributes to
the total conductance by the conductance unit $G_{0}=2e^{2}/h. $\ Figure
\ref{fig5} shows the energy spectra for zigzag and armchair BGNs with $N=46$\
and a corresponding conductance as a function of the Fermi energy at $B=0$.
Note that analytical expressions for the conductance quantization in monolayer
GNR were provided by Onipko \cite{Onipko} who showed that the conductance of
the monolayer GNRs is given by $G_{GNR}^{Z}=2(n+1/2)G_{0}$\ for the zigzag
ribbons and $G_{GNR}^{A}=nG_{0}$\ for the armchair ribbons, where the integer
number $n=0,1,2,\cdots$. Our calculations shows that the corresponding
conductance of the zigzag and armchair bilayer ribbons is given respectively by
$G_{BGN}^{Z}=2(n+1)G_{0}$\ and $G_{BGN}^{A}=nG_{0}$, see Fig. \ref{fig5}. The
minimum conductance of an ideal zigzag BGN is $2G_{0},$ whereas the minimum
conductance of an ideal metallic armchair BGN is $G_{0}.$

Note that the conductance steps are not equidistant along the energy axis. For
example, for the armchair BGN steps at $G_{0}$, $3G_{0}$, and $5G_{0}$\ are
more pronounced than those at $2G_{0}$, $4G_{0}$, and $6G_{0}$. The widths of
conductance steps are determined by the energy scale between the successive
modes in the energy spectrum, which in turn depends on the ribbon width and the
energy interval. In wider armchair BGNs, some conductance quantization steps
are poorly resolved or even unresolved which might lead to an apparent
quantization in units of $2G_{0}$.

\section{Effect of edge disorder on the conductance and localization length}

\begin{figure}[tbp]
\includegraphics[width=8cm]{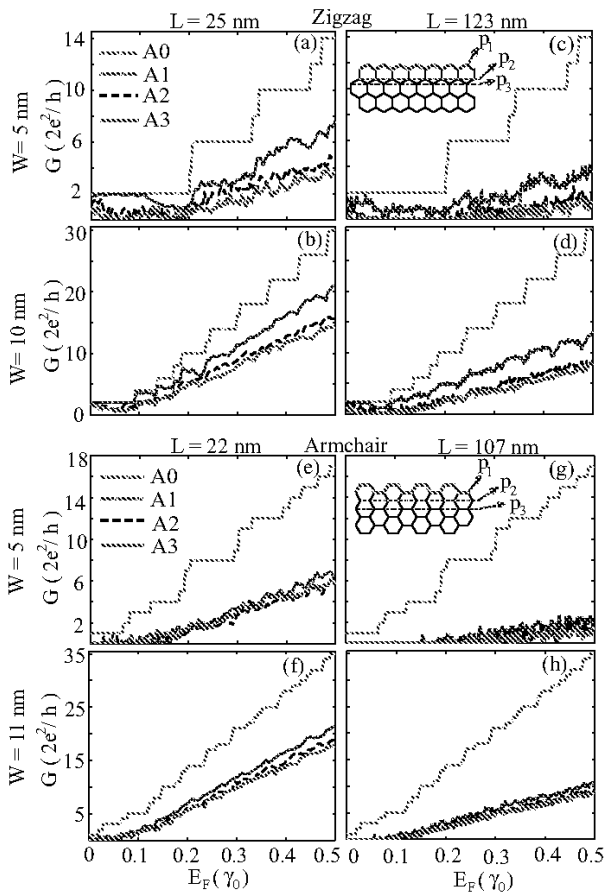} 
\caption{(Color online) The conductance as a function of the Fermi energy
with varying edge disorder for different widths and lengths. Three typical
disorder strengths are considered: $A0$: perfect edges; $A1$: $p_{1}=0.1$, $%
p_{2}=p_{3}=0$; $A2$: $p_{1}=0.3$, $p_{2}=0.2$ and $p_{3}=0.1$; $A3$: $%
p_{1}=0.5$, $p_{2}=0.4$ and $p_{3}=0.3$. The insets show the positions of
the atoms for removing with the probabilities $p_{1},p_{2},$ and $p_{3}$.
(a)-(d) the zigzag BGNs with widths $5$ nm-$10$ nm and length $25$ nm-$123$
nm. (e)-(h) the armchair BGNs with widths $5$ nm-$11$ nm and lengths $22$ nm-%
$107$ nm. }
\label{fig6}
\end{figure}

In this section we calculate the conductance of realistic BGNs with edge
defects that are created by random removal of carbon atoms from the edges of the upper
and lower layers. It was shown previously that the edge disorder
strongly affects the conductance of monolayer GNRs.\cite
{Chen2007,Han2007,Areshkin,Louis,Gunlycke,Li,Querlioz,Avouris,Evaldsson2008,Mucciolo,Basu2008,Stampfer2009}
. In particular, it has been demonstrated that even very modest edge disorder
is sufficient to strongly suppress the conductance and to induce the conduction
energy gap in the vicinity of the Dirac point and to lift any difference in the
conductance between nanoribbons of different edge geometry (i.e. zigzag and
armchair) \cite{Evaldsson2008,Mucciolo}. This was related to the pronounced
edge-disorder-induced Anderson-type localization which leads to a strongly
enhanced density of states at the edges and to blocking of conductive paths
through the ribbons\cite{Evaldsson2008}. In the present study we use the model
for the edge disorder applied previously to monolayer GNRs \cite{Evaldsson2008}
and compare the effect of the edge disorder on the conductance and the
localization length of monolayer GNRs and BGNs. We model the missing atoms by
setting the corresponding hopping matrix elements to zero. The edge roughness
is controlled by three probabilities $p_{1}$, $p_{2} $, $p_{3}$. $p_{1}$ is the
probability of a missing atom in the outermost row; $p_{i}(i=2,3)$ is the
conditional probability for a missing atom in the $i$-th row away from the edge
if at least one adjacent atom in row $(i-1)$ is missing, see the illustration in
the inset of Fig. \ref{fig6}. As graphene is known to have few bulk defects in
general \cite{Schedin} we do not consider bulk vacancies. We also disregard the
effect of hydrogen capture by dangling bonds at the edge, since it has been
shown to be of minor importance for ribbons wider than a few nanometers
\cite{Son,Barone}.

Figure \ref{fig6} shows the conductance of zigzag and armchair BGNs of
different lengths and widths as a function of the Fermi energy for four edge
disorder strengths. The conductance
of the BGNs shows the same qualitative features as the conductance of the
monolayer GNRs \cite{Evaldsson2008}. The main feature is that a relatively
small edge disorder strongly suppresses the conductance and completely destroys
the quantization steps for both zigzag and armchair BGNs. As a result, no
difference in the conductance is expected for realistic zigzag and armchair
BGNs. Armchair BGNs show a well-pronounced transport gap in the vicinity of the
charge neutrality point. As in the case of monolayer ribbons, the
zigzag BGNs are more robust to this disorder-induced suppression of the
conductance. This behavior has been related to the presence
of the edge states in the zigzag nanoribbon close to the Dirac point
\cite{Areshkin}.

\begin{figure}[tbp]
\includegraphics[width=8cm]{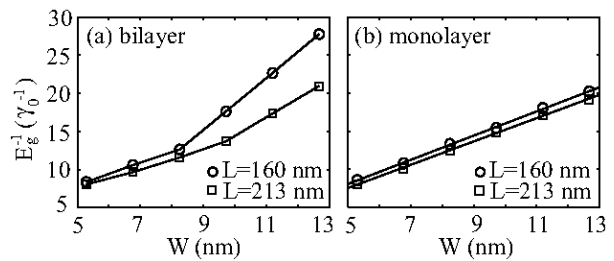} 
\caption{The inverse of the energy gap versus ribbon width for (a) armchair
BGNs and (b) armchair monolayer GNRs. The energy gap $E_g$ is defined as the
interval where $G\lesssim 10^{-3}\times 2e^2/h$. }
\label{fig7}
\end{figure}

\begin{figure}[tbp]
\includegraphics[width=8cm]{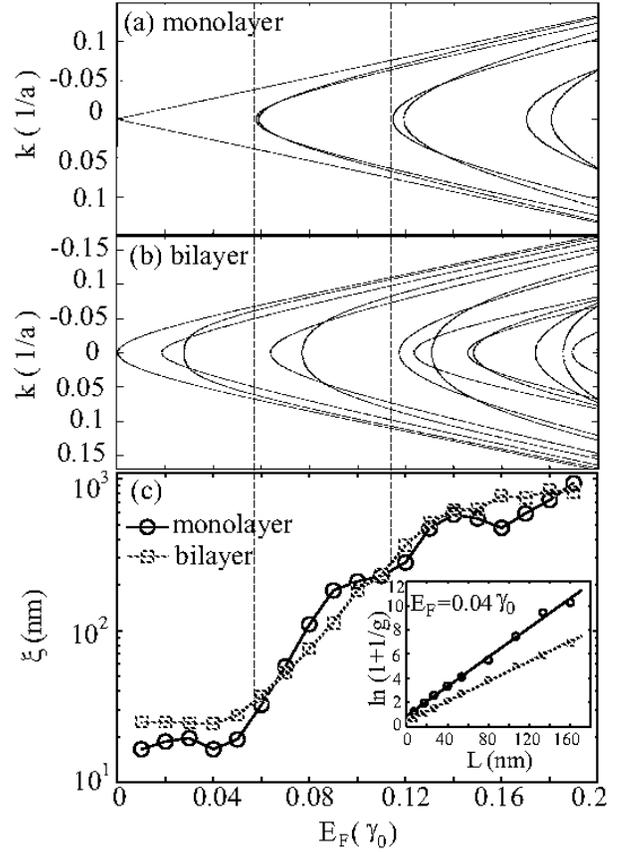} 
\caption{Bottom: The localization length $\protect\xi$ as a function of the Fermi energy
for a ribbon of the width $W=11$ nm. Inset: length dependence of $\ln (1+1/g)$. Top: The
monolayer and bilayer dispersion relations for the corresponding nanoribbons with perfect
edges.} \label{fig8}
\end{figure}

Recent experiments and calculations have shown that the transport gap in
monolayer GNRs is approximately proportional to the inverse of the ribbon width
$W$  \cite{Han2007,Louis,Querlioz,Evaldsson2008,Mucciolo}. In Fig. \ref{fig7}
we plot the inverse of the transport gap, $E_{g}^{-1},$ for the armchair BGNs
and monolayer GNRs as a function of the ribbon width $W$. While for the monolayer
ribbons $E_{g}^{GNR}\sim 1/W$, the corresponding transport gap $E_{g}^{BGN}$
for the bilayer ribbons shows a more rapid decrease as $W$ increases.
This indicates that edge-disorder induced localization of the
states is less pronounced in BGNs in comparison to GNRs. In order to elucidate this point
further, we have performed a comparative study of the localization length $
\xi $ as a function of the Fermi energy $E_F$in both systems, choosing
nanoribbons of width $W=11\,\mathrm{nm}$, corresponding to $46$ atomic sites in
 $y$-direction. The vacancy probabilities have been set to $p_{1}=0.5$, $
p_{2}=0.4$ and $p_{3}=0.3$, respectively. The nanoribbons are thus identical to
that one of Fig. \ref{fig6} (h), configuration $A3$. $\xi (E_F)$ is determined by
calculating the conductance $G$ as a function of the nanoribbon length $L$ for
an ensemble of 5000 disorder configurations for each length. As the energy
increases, the system is expected to undergo a transition from the strongly
localized regime where the conductance decays exponentially as $L$ increases, to
the Ohmic regime where $G\propto 1/L$. The conductance is expected to obey the
scaling law
\begin{equation}
\ln (1+1/g)=L/\xi ,  \label{eq4}
\end{equation}%
with the dimensionless conductance $g=G/(e^{2}/\pi \hbar )$
\cite{Anderson1980}. We find that for all energies, the simulated conductances
can be fitted with high accuracy by Eq. (\ref{eq4}), as exemplified in the
inset
of Fig. \ref{fig8}(c), with $\xi $ being the fit parameter. Fig. \ref%
{fig8} (c) shows the localization length determined that way as a function
of the Fermi energy for both monolayer and bilayer ribbons. Two regimes can
be identified, depending on whether the Fermi energy $E_{F}$ is inside or
outside the transport gap. For energies inside the transport gap, $%
|E_{F}|\lesssim E_{g},$ (with  $E_{g}\approx 0.06\gamma _{0}$ for the particular system
under study), $\xi $ is essentially independent of $E_F$ and comparable to $W$.
In this interval, $\xi$ in the BGN is roughly a
factor of $2$ larger than in the GNR. As $E_F$ increases
above the transport gap, $\xi $ increases rapidly, while the difference in
$\xi $ for the mono-and the bilayer system fluctuates around zero. We emphasize that
these fluctuations are well above the noise level, due to the large ensemble
used.

This behavior can be interpreted qualitatively as follows. Inside the
transport gap, $|E_{F}|\lesssim E_{g}$, large disorder-induced potential
barriers are present in each layer which hamper the electron transfer along
the wire. If an electron encounters such a barrier in one layer, it can
bypass it by hopping to the second layer. Because such a bypassing is
apparently not possible in a system with a single layer the electrons are
less localized in the BGNs.

Outside the transport gap, $|E_{F}|\gtrsim E_{g}$, conducting channels open up
in both layers and therefore the interlayer transfer becomes less likely
compared to the intralayer transfers. Because of this, electronic transport in
each layer is much less affected by the presence of the neighboring level which
results in similar  localization lengths for the monolayer and the bilayer
nanoribbons. The differences in $\xi $ for the mono- and the bilayer system
mentioned above can be traced to the features in the dispersion relations
corresponding to the opening of new propagating channels. A correlation between
the dispersion relation and the behavior of $\xi (E)$ is clearly seen in Fig.
\ref{fig8}. For example, the strong increase of $\xi $ between $0.06\gamma
_{0}<E<0.1\gamma _{0}$ of the monolayer is related to the occupation of the
second mode. As the energy (and therefore a number of modes) increases, the
difference of $\xi (E)$ between mono- and bilayer nanoribbons diminishes.

\begin{figure}[tbp]
\includegraphics[width=8.5cm]{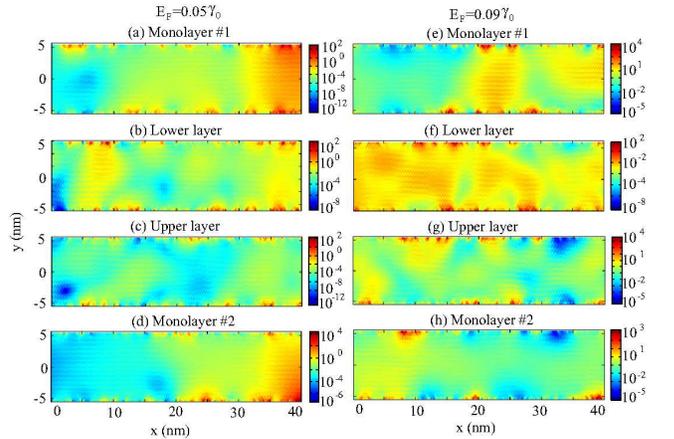} 
\caption{(Color online) The local density of states in representative parts of the edge
disorder regions. (a) and (d) are the LDOS of two independent monolayer at
$E_F=0.05\protect\gamma_0$ (i.e. inside the transport energy gap) and evolve into (b) and
(c), respectively, when the interlayer interaction is turned on. (e) and (h) are the LDOS
of two independent monolayer at $E_F=0.09\protect\gamma_0$ (i.e. outside the energy gap)
and evolve into (f) and (g), respectively, when the interlayer interaction is turned on.
} \label{fig9}
\end{figure}

An inspection of a typical local density of states (LDOS) pattern of a
representative member of the nanoribbon ensemble substantiates this
interpretation of the properties of $\xi (E)$, see Fig. \ref{fig9}. We first
generate two monolayer GNRs with the same concentration but different
configurations of edge disorder, and then couple them to form the lower and the
upper layers of a bilayer BGN. We then separately calculate the LDOS for these
three structures (i.e. for two different monolayer GNRs and one BGR). The
pronounced enhancement of the LDOS at the edges (note the logarithmic scale)
and the substantial fluctuations at larger scales indicate Anderson
localization as discussed in detail elsewhere \cite{Evaldsson2008}.

Figures \ref{fig9} (a)-(d) show the LDOS plots for $E_{F}$ lying inside the
energy gap. A comparison of the LDOS pattern shows an \emph{anticorrelation}
between the lower and upper layers of the BGN (i.e. the enhanced LDOS in one
layer is accompanied by the suppressed LDOS in the second layer).  At the
same time, there is not much correlation between LDOS patterns of the BGN
and the corresponding LDOSs of the monolayer GNRs. This is consistent with
the interpretation presented above where the \emph{interlayer} transfer is
dominant such that the electrons encountering a potential barrier (which
reduces the LDOS) in one layer, jump to the second layer and enhance the
LDOS of the neighboring sites therein.

The LDOS patterns for a representative value of $E_{F}$ above the
transport gap demonstrate that the \emph{intralayer} transfer is dominant in this
regime. Indeed, in this case a \emph{correlation} can be seen between the LDOS
patterns of the BGN and the corresponding LDOSs of the monolayer GNRs which
signifies the presence of the open propagating channels in each layer. Note
that there is also a certain \emph{correlation} even between two
layers in the BGN.   For example, the lower layer of the BGN (Fig. \ref{fig9}%
(f)) shows some additional LDOS maxima which are not present in its
constitutive monolayer $\#1$ (see Fig. \ref{fig9}(e)) but correlate to the
LDOS maxima in the second layer of the BGN (Fig. \ref{fig9} (g)). This
correlation between layers indicates that due to the coupling between the
layers the defects in the upper layer simply generate additional defects of
comparable strength in the lower layer and vice versa.

Finally we note that no attempts were made to quantify the correlations in the LDOS patterns.
We rather restricted ourselves to a visual
inspection which explain qualitatively the behavior of $\xi (E_F)$.

\section{Summary and conclusions}

We have studied the electronic structures and transport properties of
Bernal-type graphene bilayer nanoribbons using the recursive Green's
function technique. The band structures of the zigzag and armchair BGNs with
uniform perpendicular magnetic fields have been computed for the entire
energy regime of the $\pi $-bands. The calculated band structure corresponds
to two superimposed and somewhat deformed band structures of individual
monolayer GNRs. The splitting between the corresponding monolayer bands is
of the order of $\sim 0.3-0.4$ eV and is determined by the interlayer
hopping integrals $\gamma _{1}$\ and $\gamma _{3}$ providing coupling
between the layers. The conductance of the clean (disorder-free) zigzag
system is quantized as $2(n+1)G_{0}$, whereas the clean armchair system is
quantized as $nG_{0}$ with $G_{0}=2e^{2}/h$ being the conductance unit and $%
n$ an integer number. Furthermore, the effect of edge disorder, which is
highly relevant in realistic samples, has been investigated. As in the case
of monolayer ribbons, a relatively small edge disorder strongly suppress the
conductance of the BGN, introduces the transport energy gap in the vicinity
of the charge neutrality point and completely destroys the quantization
steps for both zigzag and armchair BGNs. We calculate the localization
length as a function of the Fermi energy $E_{F}$  and identify two distinct
regimes depending on whether $E_{F}$ is inside or outside the transport gap.
The localization length inside the transport gap is larger in BGNs than
in GNRs with identical edge roughness. This difference,
however, disappears at energies above the transport gap.

\begin{acknowledgments}
H.X. and T.H. acknowledge financial support from the German Academic
Exchange Service (DAAD) within the DAAD-STINT collaborative grant. I.V.Z.
acknowledge the support from the Swedish Research Council (VR) and from the
Swedish Foundation for International Cooperation in Research and Higher
Education (STINT) within the DAAD-STINT collaborative grant.
\end{acknowledgments}

\appendix

\section{Hamiltonian and coupling matrices}

\label{appdx1} In this appendix we provide explicit forms of the
hamiltonians for the ${i}$-th slices $h_{i}$ and coupling matrices $U$. $%
V_{i}^{\prime }$ denotes the potential on the site $i$ in the upper layer
and $V_{i}$ the potential on $i$-th site in the lower layer. The diagonal
blocks of the matrices correspond to the Hamiltonians of layers, and
off-diagonal blocks give the interlayer interactions.

\subsection{Zigzag edge}

\begin{equation}
h_{1}=\left(
\begin{array}{ccccc|ccccc}
V_{1}^{\prime } & \gamma _{0} & 0 & 0 & \cdots & \gamma _{1} & 0 & 0 & 0 &
\cdots \\
\gamma _{0} & V_{2}^{\prime } & 0 & 0 &  & 0 & \gamma _{3} & 0 & 0 &  \\
0 & 0 & V_{3}^{\prime } & \gamma _{0} &  & 0 & 0 & \gamma _{1} & 0 &  \\
0 & 0 & \gamma _{0} & V_{4}^{\prime } &  & 0 & 0 & 0 & \gamma _{3} &  \\
\vdots &  &  &  & \ddots & \vdots &  &  &  & \ddots \\ \hline
\gamma _{1} & 0 & 0 & 0 & \cdots & V_{1} & 0 & 0 & 0 & \cdots \\
0 & \gamma _{3} & 0 & 0 &  & 0 & V_{2} & \gamma _{0} & 0 &  \\
0 & 0 & \gamma _{1} & 0 &  & 0 & \gamma _{0} & V_{3} & 0 &  \\
0 & 0 & 0 & \gamma _{3} &  & 0 & 0 & 0 & V_{4} &  \\
\vdots &  &  &  & \ddots & \vdots &  &  &  & \ddots \\
&  &  &  &  &  &  &  &  &
\end{array}%
\right)  \label{A1}
\end{equation}

\begin{equation}
h_{2}=\left(
\begin{array}{ccccc|ccccc}
V_{1}^{\prime } & 0 & 0 & 0 & \cdots & \gamma _{3} & 0 & 0 & 0 & \cdots \\
0 & V_{2}^{\prime } & \gamma _{0} & 0 &  & 0 & \gamma _{1} & 0 & 0 &  \\
0 & \gamma _{0} & V_{3}^{\prime } & 0 &  & 0 & 0 & \gamma _{3} & 0 &  \\
0 & 0 & 0 & V_{4}^{\prime } &  & 0 & 0 & 0 & \gamma _{1} &  \\
\vdots &  &  &  & \ddots & \vdots &  &  &  & \ddots \\ \hline
\gamma _{3} & 0 & 0 & 0 & \cdots & V_{1} & \gamma _{0} & 0 & 0 & \cdots \\
0 & \gamma _{1} & 0 & 0 &  & \gamma _{0} & V_{2} & 0 & 0 &  \\
0 & 0 & \gamma _{3} & 0 &  & 0 & 0 & V_{3} & \gamma _{0} &  \\
0 & 0 & 0 & \gamma _{1} &  & 0 & 0 & \gamma _{0} & V_{4} &  \\
\vdots &  &  &  & \ddots & \vdots &  &  &  & \ddots \\
&  &  &  &  &  &  &  &  &
\end{array}%
\right)  \label{A2}
\end{equation}

\begin{equation}
U^{12}=(U^{21})^{+}=\left(
\begin{array}{ccccc|ccccc}
\gamma _{0} & 0 & 0 & 0 & \cdots & 0 & 0 & 0 & 0 & \cdots \\
0 & \gamma _{0} & 0 & 0 &  & \gamma _{3} & 0 & 0 & 0 &  \\
0 & 0 & \gamma _{0} & 0 &  & 0 & 0 & 0 & 0 &  \\
0 & 0 & 0 & \gamma _{0} &  & 0 & 0 & \gamma _{3} & 0 &  \\
\vdots &  &  &  & \ddots &  &  &  &  & \ddots \\ \hline
0 & 0 & 0 & 0 & \cdots & \gamma _{0} & 0 & 0 & 0 & \cdots \\
0 & 0 & \gamma _{3} & 0 &  & 0 & \gamma _{0} & 0 & 0 &  \\
0 & 0 & 0 & 0 &  & 0 & 0 & \gamma _{0} & 0 &  \\
0 & 0 & 0 & 0 &  & 0 & 0 & 0 & \gamma _{0} &  \\
\vdots &  &  &  & \ddots & \vdots &  &  &  & \ddots \\
&  &  &  &  &  &  &  &  &
\end{array}%
\right)  \label{A3}
\end{equation}

\subsection{Armchair edge}

\begin{equation}
h_{1\sim 4}=\left(
\begin{array}{ccccc|ccccc}
V_{1}^{\prime } & 0 & 0 & 0 & \cdots &  &  &  &  &  \\
0 & V_{2}^{\prime } & 0 & 0 &  &  &  &  &  &  \\
0 & 0 & V_{3}^{\prime } & 0 &  &  &  & 0 &  &  \\
0 & 0 & 0 & V_{4}^{\prime } &  &  &  &  &  &  \\
\vdots &  &  &  & \ddots &  &  &  &  &  \\ \hline
&  &  &  &  & V_{1} & 0 & 0 & 0 & \cdots \\
&  &  &  &  & 0 & V_{2} & 0 & 0 &  \\
&  & 0 &  &  & 0 & 0 & V_{3} & 0 &  \\
&  &  &  &  & 0 & 0 & 0 & V_{4} &  \\
&  &  &  &  & \vdots &  &  &  & \ddots \\
&  &  &  &  &  &  &  &  &
\end{array}%
\right)  \label{A4}
\end{equation}

\begin{equation}
U^{01}=(U^{10})^{+}=\left(
\begin{array}{ccccc|ccccc}
\gamma _{0} & 0 & 0 & 0 & \cdots &  &  &  &  &  \\
0 & \gamma _{0} & 0 & 0 &  &  &  &  &  &  \\
0 & 0 & \gamma _{0} & 0 &  &  &  & 0 &  &  \\
0 & 0 & 0 & \gamma _{0} &  &  &  &  &  &  \\
\vdots &  &  &  & \ddots &  &  &  &  &  \\ \hline
\gamma _{3} & 0 & 0 & 0 & \cdots & \gamma _{0} & 0 & 0 & 0 & \cdots \\
\gamma _{3} & \gamma _{3} & 0 & 0 &  & 0 & \gamma _{0} & 0 & 0 &  \\
0 & \gamma _{3} & \gamma _{3} & 0 &  & 0 & 0 & \gamma _{0} & 0 &  \\
0 & 0 & \gamma _{3} & \gamma _{3} &  & 0 & 0 & 0 & \gamma _{0} &  \\
\vdots &  &  &  & \ddots & \vdots &  &  &  & \ddots \\
&  &  &  &  &  &  &  &  &
\end{array}%
\right)  \label{A5}
\end{equation}

\begin{equation}
U^{12}=(U^{21})^{+}=\left(
\begin{array}{ccccc|ccccc}
\gamma _{0} & \gamma _{0} & 0 & 0 & \cdots & \gamma _{3} & 0 & 0 & 0 & \cdots
\\
0 & \gamma _{0} & \gamma _{0} & 0 &  & 0 & \gamma _{3} & 0 & 0 &  \\
0 & 0 & \gamma _{0} & \gamma _{0} &  & 0 & 0 & \gamma _{3} & 0 &  \\
0 & 0 & 0 & \gamma _{0} &  & 0 & 0 & 0 & \gamma _{3} &  \\
\vdots &  &  &  & \ddots & \vdots &  &  &  & \ddots \\ \hline
\gamma _{1} & 0 & 0 & 0 & \cdots & \gamma _{0} & 0 & 0 & 0 & \cdots \\
0 & \gamma _{1} & 0 & 0 &  & \gamma _{0} & \gamma _{0} & 0 & 0 &  \\
0 & 0 & \gamma _{1} & 0 &  & 0 & \gamma _{0} & \gamma _{0} & 0 &  \\
0 & 0 & 0 & \gamma _{1} &  & 0 & 0 & \gamma _{0} & \gamma _{0} &  \\
\vdots &  &  &  & \ddots & \vdots &  &  &  & \ddots \\
&  &  &  &  &  &  &  &  &
\end{array}%
\right)  \label{A6}
\end{equation}

\begin{equation}
U^{23}=(U^{32})^{+}=\left(
\begin{array}{ccccc|ccccc}
\gamma _{0} & 0 & 0 & 0 & \cdots &  &  &  &  &  \\
0 & \gamma _{0} & 0 & 0 &  &  &  &  &  &  \\
0 & 0 & \gamma _{0} & 0 &  &  &  & 0 &  &  \\
0 & 0 & 0 & \gamma _{0} &  &  &  &  &  &  \\
\vdots &  &  &  & \ddots &  &  &  &  &  \\ \hline
\gamma _{3} & \gamma _{3} & 0 & 0 & \cdots & \gamma _{0} & 0 & 0 & 0 & \cdots
\\
0 & \gamma _{3} & \gamma _{3} & 0 &  & 0 & \gamma _{0} & 0 & 0 &  \\
0 & 0 & \gamma _{3} & \gamma _{3} &  & 0 & 0 & \gamma _{0} & 0 &  \\
0 & 0 & 0 & \gamma _{3} &  & 0 & 0 & 0 & \gamma _{0} &  \\
\vdots &  &  &  & \ddots & \vdots &  &  &  & \ddots \\
&  &  &  &  &  &  &  &  &
\end{array}%
\right)  \label{A7}
\end{equation}

\begin{equation}
U^{34}=(U^{43})^{+}=\left(
\begin{array}{ccccc|ccccc}
\gamma _{0} & 0 & 0 & 0 & \cdots & \gamma _{3} & 0 & 0 & 0 & \cdots \\
\gamma _{0} & \gamma _{0} & 0 & 0 &  & 0 & \gamma _{3} & 0 & 0 &  \\
0 & \gamma _{0} & \gamma _{0} & 0 &  & 0 & 0 & \gamma _{3} & 0 &  \\
0 & 0 & \gamma _{0} & \gamma _{0} &  & 0 & 0 & 0 & \gamma _{3} &  \\
\vdots &  &  &  & \ddots & \vdots &  &  &  & \ddots \\ \hline
\gamma _{1} & 0 & 0 & 0 & \cdots & \gamma _{0} & \gamma _{0} & 0 & 0 & \cdots
\\
0 & \gamma _{1} & 0 & 0 &  & 0 & \gamma _{0} & \gamma _{0} & 0 &  \\
0 & 0 & \gamma _{1} & 0 &  & 0 & 0 & \gamma _{0} & \gamma _{0} &  \\
0 & 0 & 0 & \gamma _{1} &  & 0 & 0 & 0 & \gamma _{0} &  \\
\vdots &  &  &  & \ddots & \vdots &  &  &  & \ddots \\
&  &  &  &  &  &  &  &  &
\end{array}%
\right)  \label{A8}
\end{equation}

\section{Formalism}

\label{appdx2} In this Appendix we provide main formulas used for calculation
of the dispersion relations, Bloch states and their velocities and the
transmission and reflection amplitudes for the scattering problem for bilayer
graphene ribbons. All these formulas represent a straighforward generalization
of the corresponding formular derived in Ref. [\onlinecite{Xu2008}] for
monolayer ribbons.

The band structure is computed by solving a eigenvalue of the form
\begin{equation}
T_{1}^{-1}T_{2}\left(
\begin{array}{c}
\psi _{1} \\
\psi _{0}%
\end{array}%
\right) =e^{ikM}\left(
\begin{array}{c}
\psi _{1} \\
\psi _{0}%
\end{array}%
\right)  \label{Bloch_eigenequation}
\end{equation}%
with $M$ being the periodicity ($M=2$ and $4$ for the zigzag and armchair
BGRs respectively),
\begin{equation*}
T_{1}=%
\begin{pmatrix}
-G_{\mathrm{cell}}^{1,M}U^{M,M+1} & \;0 \\
-G_{\mathrm{cell}}^{M,M}U^{M,M+1} & \;I%
\end{pmatrix}%
,\;T_{2}=%
\begin{pmatrix}
-I & \;G_{\mathrm{cell}}^{1,1}U^{1,0} \\
0 & \;G_{\mathrm{cell}}^{M,1}U^{1,0}%
\end{pmatrix}%
\end{equation*}%
\newline
$I$ being the unitary matrix.

This eigenvalue problem Eq. (\ref{Bloch_eigenequation}) gives the
eigenfunctions $\psi ^{\alpha }$, $1\leq \alpha \leq 2N$ and the set of
Bloch eigenvectors $\left\{ k_{\alpha }\right\} $ which includes both
propagating and evanescent states. The latter can be easily identified by a
non-zero imaginary part. In order to separate right- and left-propagating
states, $k_{\alpha }^{+}$ and $k_{a}^{-},$ we compute the group velocities
of the Bloch states $v_{\alpha }=\frac{\partial E}{\partial k_{\alpha }}$,
whose signs determine the direction of propagation (`+' stands for the
right-propagating and `-' for the left propagating states).

Starting from the Schr\"{o}dinger equation and using a definition of the
group velocity, we obtain the group velocity for graphene nanoribbons
\begin{equation}
v=\frac{1}{M}\sum_{i=1}^{M}\frac{\partial }{\partial k}\left[ \frac{\langle
\psi _{i}|H|\psi \rangle }{|\varphi _{i}|^{2}}\right]
\end{equation}%
where the summation is performed over all slices of the unit cell, and
\begin{equation}
\varphi _{i}=(\varphi _{i,1};...;\varphi _{i,2N})^{T}  \label{phi_vector}
\end{equation}%
is a vector composed of the matrix elements $\varphi _{i,j}=$ $\langle
0a_{i,j}|\varphi \rangle $ (Note that vectors $\varphi _{i}$ can be obtained
from $\psi _{i}$ via the relation $\psi _{i}=e^{iki}\varphi _{i}$).

To account for the effects of the leads, one needs to calculate the surface
Green's function. The latter equations can be used for determination of $%
\Gamma _{r}$,%
\begin{equation}
\Gamma _{r}U^{1,0}=\Psi _{1}\Psi _{0}^{-1},  \label{Gamma_r}
\end{equation}%
where $\Psi _{1}$ and $\Psi _{0}$ are the square matrixes composed of the
matrix-columns $\psi _{1}^{\alpha }$ and $\psi _{0}^{\alpha },$ ($1\leq
\alpha \leq 2N),$ Eq. (\ref{Bloch_eigenequation}), i.e. $\Psi _{1}=(\psi
_{1}^{1},...,\psi _{1}^{2N});$ $\Psi _{0}=(\psi _{0}^{1},...,\psi
_{0}^{2N}). $ The expression for the left surface Greens function $\Gamma
_{l}$ (i.e. the surface function of the semiinfinite ribbon open to the
right) is derived in a similar fashion,%
\begin{equation}
\Gamma _{l}U^{M,M+1}=\Psi _{M}\Psi _{M+1}^{-1},  \label{Gamma_l}
\end{equation}%
where the matrixes $\Psi _{M}$ and $\Psi _{M+1}$ are defined in a similar
way as $\Psi _{1}$ and $\Psi _{0}$ above.

The transmission and reflection amplitudes is calculated using the
expressions:
\begin{equation}
\Phi _{1}T=-G^{L,0}(U^{0,1}\Psi _{1}-\Gamma _{l}^{-1}\Phi _{0}).  \label{T}
\\
\end{equation}%
\begin{equation}
\Phi _{0}R=-G^{0,0}(U^{0,1}\Psi _{1}-\Gamma _{l}^{-1}\Phi _{0})-\Phi _{0}.
\end{equation}%
The matrices $T$ and $R$ with the dimension 2$N\times N_{\mathrm{prop}}$ are
composed of the transmission and reflection amplitudes $t_{\beta \alpha }$
and $r_{\beta \alpha }$, (with $N_{\mathrm{prop}}$ being the number of
propagating modes in the leads). The Green's functions $G^{M+1,0}$ and $%
G^{0,0}$ are obtained from the standard recursive technique based on the
Dyson's equation \cite{Ferry}. The transmission and reflection are related
to their amplitudes by $(T)_{\beta \alpha }=v_{\beta }^{+}/v_{\alpha
}^{+}|t_{\beta \alpha }|^{2}$ and $(R)_{\beta \alpha }=v_{\beta
}^{-}/v_{\alpha }^{+}|r_{\beta \alpha }|^{2}$.

\end{document}